\documentclass[%
prx,
 amsmath,amssymb,
 aps,
twocolumn,
longbibliography,
showkeys
]{revtex4-1}

\usepackage{graphicx}
\usepackage{epstopdf}
\usepackage{dcolumn}
\usepackage{bm}
\usepackage{color}
\usepackage[mathlines]{lineno}
\usepackage[breaklinks=true,colorlinks=true,linkcolor=blue,urlcolor=blue,citecolor=blue]{hyperref}
\usepackage{ragged2e}
\usepackage{enumitem}
\usepackage{bibunits}
\usepackage{amsthm}
\usepackage{diagbox}

\newcommand\Reds[1]{{\color{black}#1}}

\begin{document}
\title{Constructive agents nullify the ability of destructive agents to foster cooperation in public goods games}

\author{Yuting Dong$^1$}
\author{Zhixue He$^{1,2}$}
\author{Chen Shen$^3$}
\email{steven\_shen91@hotmail.com}
\author{Lei Shi$^1$}
\email{shi\_lei65@hotmail.com}
\author{Jun Tanimoto$^{2,3}$}

\affiliation{
\vspace{2mm}
\mbox{1. School of Statistics and Mathematics, Yunnan University of Finance and Economics, Kunming, 650221, China}
\mbox{2. Interdisciplinary Graduate School of Engineering Sciences, Kyushu University, Fukuoka, 816-8580, Japan}
\mbox{3. Faculty of Engineering Sciences, Kyushu University, Kasuga-koen, Kasuga-shi, Fukuoka 816-8580, Japan}
}

\date{\today}

\begin{abstract}
Existing studies have revealed a paradoxical phenomenon in public goods games, wherein destructive agents, harming both cooperators and defectors, can unexpectedly bolster cooperation. Building upon this intriguing premise, our paper introduces a novel concept: constructive agents, which confer additional benefits to both cooperators and defectors. We investigate the impact of these agents on cooperation dynamics within the framework of public goods games. Employing replicator dynamics, we find that unlike destructive agents, the mere presence of constructive agents does not significantly alter the defective equilibrium. However, when the benefits from constructive agents are outweighed by the damage inflicted by destructive agents, the addition of constructive agents does not affect the ability of destructive agents to sustain cooperation. In this scenario, cooperators can be maintained through a cyclic dominance between cooperators, defectors, and destructive agents, with constructive agents adding complexity but not fundamentally changing the equilibrium. Conversely, if the benefits from constructive agents surpass the harm caused by destructive agents, the presence of constructive agents nullifies the ability of destructive agents to foster cooperation. Our results highlight the nuanced role of constructive agents in cooperation dynamics, emphasizing the necessity of carefully assessing incentive balances when encouraging cooperation.
\end{abstract}

\keywords{Evolutionary game theory; Constructive agents; Destructive agents; Public goods game}

\maketitle

\section{Introduction}  
Cooperation can ensure the maximization of collective interests, but it is also vulnerable to exploitation by ``free-riders" who do not bear the costs of cooperation \cite{axelrod1981evolution,riolo2001evolution,archetti2012game}. To explore the emergence and maintenance of cooperation, numerous studies rooted in the framework of evolutionary game theory have revealed the mechanisms that support the evolution of cooperation in individual interactions \cite{weibull1997evolutionary,nowak2011supercooperators,rand2013human,guo2024cooperation,nowak2006five}. These include direct reciprocity, established through repeated interactions \cite{schmid2021unified}; indirect reciprocity involving information transfer \cite{leimar2001evolution}; and network reciprocity arising from interaction structures \cite{nowak2006five}. However, in one-shot games where individuals interact only once, without the possibility of memory or behavioral information transfer, these reciprocity mechanisms are absent, making the maintenance of cooperation still a challenge \cite{riolo2001evolution}. To tackle cooperation challenges in one-shot games, researchers have incorporated social mechanisms that allow individuals to engage in additional decision-making stages after contributing to public goods in the public goods game (PGG), unveiling pathways for establishing cooperation \cite{nowak2011supercooperators,fehr2000cooperation,balliet2011reward,szolnoki2010reward}. These include allowing individuals to impose costly punishments \cite{fehr2000cooperation,li2018punishment,szolnoki2017second} and rewards \cite{andreoni2003carrot,chen2015first} on others based on contribution outcomes, deciding whether to participate in the game before contributing to public goods (or called ``loner'' strategy) \cite{szabo2002phase,szabo2002evolutionary}, or exiting from the interaction early for a fixed payoff \cite{shen2021exit,li2024granting}. From these game interactions, different behaviors can be observed. The framework of social value orientation analyzes and summarizes these behaviors by exploring how individuals balance their own interests with those of others\cite{bogaert2008social,murphy2011measuring}, aiding in understanding cooperation issues in social dilemmas from a perspective of behavioral value orientation.

Recently, attention has been drawn to the impact of sadistic-orientation behavior on the cooperation dynamics, known as destructive behavior (or called ``joker'' strategy) \cite{arenas2011joker,requejo2012stability,khatun2024stability}. These ``bad guys'' not only opt out of contributing to public goods, but also engage in antisocial behaviors that harm the interests of others. Counterintuitively, the presence of these destructive agents can maintain cooperation through a cycle of dominance involving cooperation, defection, and destructive strategies \cite{arenas2011joker}, and these individuals' positive effects on cooperation remain robust in both infinite and finite populations \cite{requejo2012stability}. However, in contrast to this sadistic-orientation behavior, altruistic-orientation behavior is more commonly observed in behavioral experiments \cite{bogaert2008social,grund2013natural}. For instance, some individuals exhibit a preference for engaging in altruistic rewarding behavior, even at a personal cost \cite{choi2013strategic}. The complexity of human behavior leads to diversity in actions and strategies in interactions \cite{bogaert2008social,murphy2011measuring}. Taking this into account, research further explores the evolutionary dynamics of coexisting altruistic and antisocial behaviors \cite{gneezy2012conflict}. Interestingly, Szolnoki et al. \cite{szolnoki2017second} demonstrated that antisocial punishment (punishing those who contribute to the public good) can unexpectedly enhance the effectiveness of prosocial punishment  (punishing free-riders) in structured populations, challenging conventional perspectives on prosocial/antisocial behaviors and revealing new pathways for establishing cooperation. Importantly, the dynamics created by diverse interaction strategies provide a crucial perspective for a deeper understanding of the emergence of cooperation in real-life scenarios, which lie in the complexity of human behavior \cite{santos2012role,perc2008social}.

Inspired by this, we introduce a new concept—constructive strategy, incorporating this altruistic behavior into the PGG model involving cooperation, defection, and destructive strategies. This approach enables us to  expand current research in solving the cooperation issues by exploring the interconnected diverse behaviors of cooperation dynamics. In contrast to the destructive agents, constructive agents withdraw from game interactions but provide additional costly benefits to participants without concern for their own gains or losses. Utilizing replicator dynamics in an infinite population \cite{smith1982evolution}, our research reveals that, although constructive agents alone do not alter evolutionary dynamics, \Reds{when the benefits from constructive agents are outweighed by the damage inflicted by destructive agents, the addition of constructive agents does not affect the ability of destructive agents to sustain cooperation.}
However, when the benefits provided by constructive agents surpass the harm caused by destructive agents, it can result in negative effects that diminish the capacity of destructive agents to promote cooperation. The role of constructive individuals emphasizes the balancing impact of positive and negative incentive factors in evolutionary game dynamics, thereby enhancing the understanding of the incentive forces shaping cooperation.

\section{Model} 

To gain a deep understanding of how constructive agents influence the cooperation dynamics, we first examine a three-strategy model that includes cooperation, defection, and construction within the PGG. We then explore a four-strategy PGG model incorporating cooperation, defection, construction, and destruction to study the evolutionary dynamics of cooperation in the presence of both constructive agents and destructive agents.

\subsection{Public goods game with constructive agents}

In a classic PGG involving $N$ players, a cooperator contributes $c$ to the public pool (simplify without loss of generality, set $c=1$), while a defector contributes nothing. The total amount in the public pool is multiplied by a synergy factor $r$ and then evenly distributed among all participants. Defection is the only Nash equilibrium strategy when the synergy factor for cooperators meets $1< r < N$ \cite{archetti2012game}. When constructive agents are introduced, these individuals neither contribute to nor benefit from the public pool. Instead, they offer an additional benefit of $d_{2} > 0$ to the participants (i.e., cooperators and defectors). Let the number of cooperators among the other individuals in the group be $N_{C}$, and the number of constructive agents be $N_{CA}$, satisfying $0 \leq N_{C} + N_{CA} \leq \Reds{N-1}$, the payoff for a focal individual adopting cooperation ($C$), defection ($D$), and constructive agents strategies ($CA$) can be respectively expressed as follows:
\begin{equation}
\begin{gathered}
\pi_{C} =\frac{r (N_{C} + 1)  +d_{2} N_{CA}}{S_{1}}-1,\\
\pi_{D} =\frac{r N_{C} +d_{2} N_{CA}}{S_{1}},\\
\pi_{CA} =0,
\end{gathered}
\label{eq1}
\end{equation}
where $S_{1} = N - N_{CA}$ is the number of non-constructive agents in the group. 

Consider an infinitely large and well-mixed population consisting of cooperators, defectors, and constructive agents, with proportions $x$, $y$, $z$, respectively (satisfy $x+y+z=1$ and $0\leq x,y,z \leq 1 $). $N$ individuals are randomly selected from the population to form group and play the PGG. The expected payoffs for the $C$, $D$, and $CA$ strategies can be calculated as follows:
\begin{equation}
\begin{aligned}
P_{C} & = \sum_{N_{CA}=0}^{N-1} \sum_{N_{C}=0}^{N-1-N_{CA}}\binom{N-1}{N_{CA}}\binom{N-1-N_{CA}}{N_{C}} \\
& \quad \cdot x^{N_{C}} \cdot z^{N_{CA}} \cdot(1-x-z)^{N-1-N_{C}-N_{CA}} \cdot \pi_{C} \\
& = r \frac{x}{1-z}\left(1-\frac{1-(z)^{N}}{N(1-z)}\right)+d_{2}\left(\frac{\left(1-(z)^{N}\right)}{(1-z)}-1\right)\\
& \quad +\frac{r\left[1-(z)^{N}\right]}{N(1-z)}-1 ,
\\
P_{D} &= \sum_{N_{CA}=0}^{N-1} \sum_{N_{C}=0}^{N-1-N_{CA}}\binom{N-1}{N_{CA}}\binom{N-1-N_{CA}}{N_{C}} \\
& \quad \cdot x^{N_{C}} \cdot z^{N_{CA}} \cdot(1-x-z)^{N-1-N_{C}-N_{CA}} \cdot \pi_{D} \\
& = r \frac{x}{1-z}\left(1-\frac{1-(z)^{N}}{N(1-z)}\right)+d_{2}\left(\frac{\left(1-(z)^{N}\right)}{(1-z)}-1\right) ,
\\
P_{CA} &= 0 .
\end{aligned}
\label{eq3}
\end{equation}
By utilizing replicator dynamics \cite{smith1982evolution}, and taking into account the existence of mutations where the probability of an individual mutating to another type is $\mu$ ($\mu \ll 1$), the evolutionary dynamics of strategies in this population can be write as:
\begin{equation}
\begin{cases}
\dot{x}=x\left(P_{C}-\bar{P}\right)+\mu(1-3 x)\\
\dot{y}=y\left(P_{D}-\bar{P}\right)+\mu(1-3 y)\\
\dot{z}=z\left(P_{CA}-\bar{P}\right)+\mu(1-3 z)
\end{cases},
\label{eq2}
\end{equation}
where $\bar{P}=xP_{C}+yP_{D}+zP_{CA}$ is the expected payoff of the population.

\subsection{Public goods game with constructive agents and destructive agents}

When considering the four-strategy model, destructive agents (denoted as $DA$), like $CA$, do not participate in public pool allocation and investment, but indiscriminately reduce the payoffs of cooperators and defectors by $d_1$ \cite{arenas2011joker}. Let $N_{DA}$ represent the number of destructive agents, and $S_{2}$ denote the individuals who are neither constructive agents nor destructive agents, i.e., $S_{2}=N-N_{CA}-N_{DA}$. Similar to Eq. (\ref{eq1}), the payoffs of focal players adopting $C$, $D$, $CA$, and $DA$ strategies within a group can be expressed as follows:
\begin{equation}
\begin{gathered}
\pi_{C}=\frac{r\left(N_{C}+1\right)+d_{2} N_{CA}-d_{1} N_{DA}}{S_{2}}-1,\\
\pi_{D}=\frac{r N_{C}+d_{2} N_{CA}-d_{1} N_{DA}}{S_{2}},\\
\pi_{DA}=0,\\
\pi_{CA}=0.
\end{gathered}
\label{eq4}
\end{equation}
Let the proportion of $C$, $D$, $CA$ and $DA$ in an infinitely large and well-mixed population be $x$, $y$, $z$, and $w$, respectively (satisfy $x+y+z+w=1$ and $0\leq x,y,z,w \leq 1 $). The expected payoffs for these four types of agents can be calculated as follows:
\begin{equation}
\begin{aligned}
P_{C} &= \sum_{N_{CA}=0}^{N-1} \sum_{N_{D A}=0}^{N-1-N_{C A}} \sum_{N_{C}=0}^{N-1-N_{CA}-N_{D A}}\binom{N-1}{N_{CA}} \\
& \quad \binom{N-1-N_{CA}}{N_{DA}}\binom{N-1-N_{DA}-N_{CA}}{N_{C}} \\
& \quad \cdot x^{N_{C}} \cdot z^{N_{CA}} \cdot w^{N_{DA}} \\
& \quad \cdot(1-x-z-w)^{N-1-N_{C}-N_{CA}-N_{DA}} \cdot \pi_{C} \\
& = r \frac{x}{1-z-w}\left(1-\frac{1-(z+w)^{N}}{N(1-z-w)}\right)\\
& \quad -d_{1} \frac{w\left(1-(z+w)^{N-1}\right)}{(1-z-w)} \quad +d_{2} \frac{z\left(1-(z+w)^{N-1}\right)}{(1-z-w)}\\
& \quad +\frac{r}{N} \frac{\left[1-(z+w)^{N}\right]}{(1-z-w)}-1,\\
\end{aligned}
\end{equation}
\begin{equation}
\begin{aligned}
P_{D} &= \sum_{N_{CA}=0}^{N-1} \sum_{N_{D A}=0}^{N-1-N_{C A}} \sum_{N_{C}=0}^{N-1-N_{CA}-N_{D A}}\binom{N-1}{N_{CA}} \\
& \quad \binom{N-1-N_{CA}}{N_{DA}}\binom{N-1-N_{DA}-N_{CA}}{N_{C}} \\
& \quad \cdot x^{N_{C}} \cdot z^{N_{CA}} \cdot w^{N_{DA}} \\
& \quad \cdot(1-x-z-w)^{N-1-N_{C}-N_{CA}-N_{DA}} \cdot \pi_{D} \\
& = r \frac{x}{1-z-w}\left(1-\frac{1-(z+w)^{N}}{N(1-z-w)}\right)\\
& \quad -d_{1} \frac{w\left(1-(z+w)^{N-1}\right)}{(1-z-w)}  \quad +d_{2} \frac{z\left(1-(z+w)^{N-1}\right)}{(1-z-w)}\\
\end{aligned}
\end{equation}
\begin{equation}
\begin{aligned}
P_{CA} & = 0 , \\
\end{aligned}
\end{equation}
\begin{equation}
\begin{aligned}
P_{DA} & = 0 .
\end{aligned}
\end{equation}
Similar to Eq. (\ref{eq2}), the evolutionary dynamics in this population can be described as: 
\begin{equation}
   \begin{cases}
   \dot{x}=x\left(P_{C}-\bar{P}\right)+\mu(1-4 x),\\
   \dot{y}=y\left(P_{D}-\bar{P}\right)+\mu(1-4 y),\\
   \dot{z}=z\left(P_{CA}-\bar{P}\right)+\mu(1-4 z),\\
   \dot{w}=w\left(P_{DA}-\bar{P}\right)+\mu(1-4 w),
   \end{cases}
	\label{eq5}
 \end{equation}
where $\bar{P}=xP_{C}+yP_{D}+zP_{CA}+wP_{DA}$ is the average expected payoff of the population.

\section{Results} 

Before presenting our research results, we review the dynamics of the destructive agents in the traditional public goods game \cite{arenas2011joker,requejo2012stability}. In a population of size \Reds{$M$}, it is assumed that all individuals hold the same strategy. The invasion analysis in ref.~\cite{arenas2011joker} indicates that the destructive agents can lead to three patterns of invasion under the parameter conditions $1 < r < r_{\max} = N(M-1)/(M-N)$ and $d_{1} > 0$. (i) In the region of $r > 1 + (N-1) d_{1}$, destructive agents support the emergence of cooperators trough a rock-paper-scissors (RPS) cyclic dominance emerges. (ii) In the region of $1+d_{1} /(M-1)< r < 1 + (N-1) d_{1}$, a bistability arises between the destructive agents-cooperators. Here, the destructive agents can prevail over the defector, while the defector  dominates the cooperator. However, neither the destructive agents nor the cooperator can invade each other. (iii) In the region of $r < 1 + d_{1}/(M-1)$, the destructive agents dominates in population. 

\begin{figure}[!ht]
\centering
\includegraphics[width=1\linewidth]{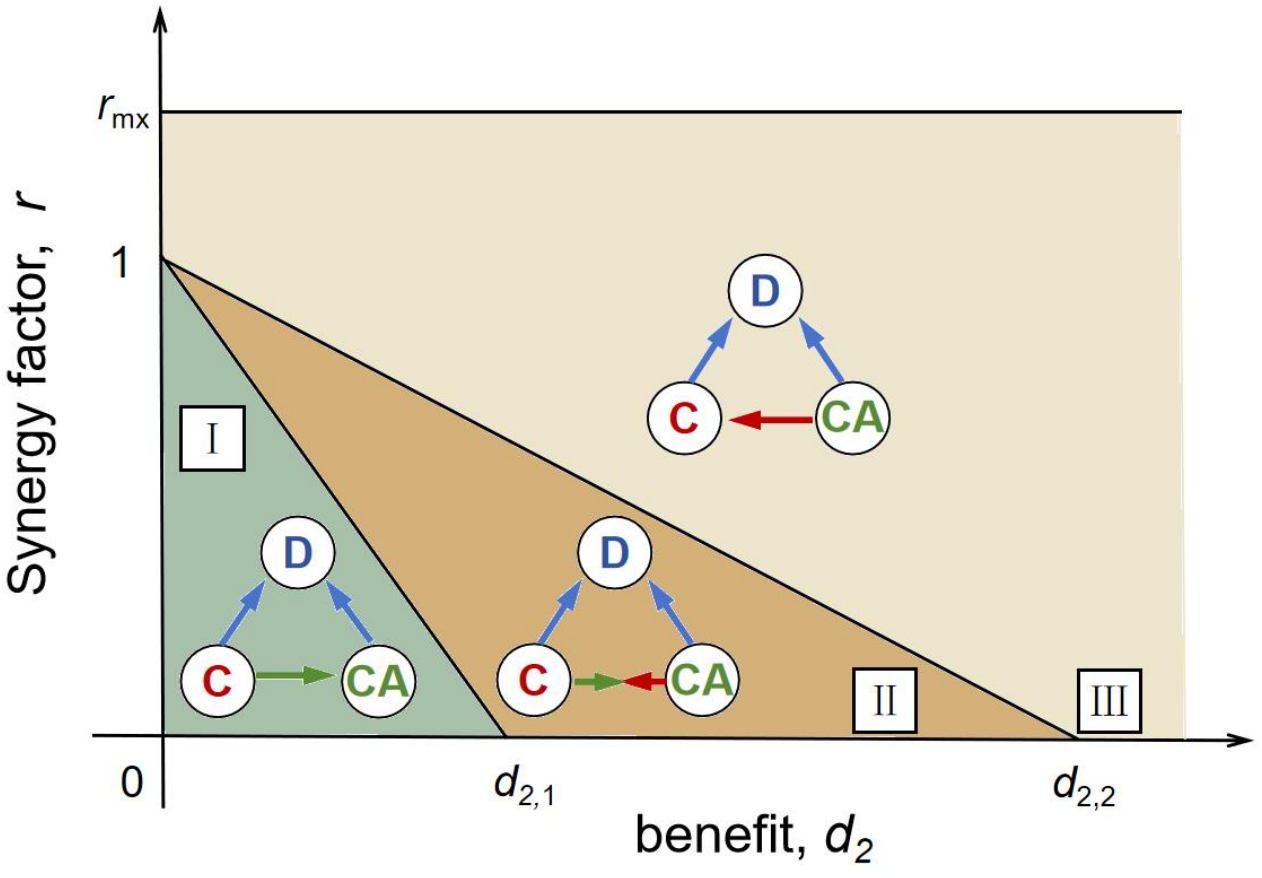}
\caption{\label{fig1}
Three patterns of invasion in the public goods games involving cooperators ($C$), defectors ($D$) and constructive agents ($CA$) under parameter phase $r$-$d_2$. Region I (light green area) satisfies the conditions $r < 1-(N-1) d_{2}$, cooperators are invaded by constructive agents, while both cooperators and constructive agents are invaded by defectors. Region II (light brown area) satisfies the conditions $1-(N-1)d_{2}< r < 1-d_{2}/(M-1)$, cooperators are invaded by constructive agents, while both cooperators and constructive agents are invaded by defectors. Region III (light yellow area) satisfies the conditions $r > 1-d_{2}/(M-1)$, constructive agents are invaded by cooperators, while both cooperators and constructive agents are invaded by defectors. All these cases are applicable under $N > 1$ and $0<r<r_{\max} =N(M-1)/(M-N)$. Critical points are $d_{2,1} = 1/(N-1)$, $d_{2,2} = M-1$.
}
\end{figure}

\begin{figure}[!ht]
\centering
\includegraphics[width=1\linewidth]{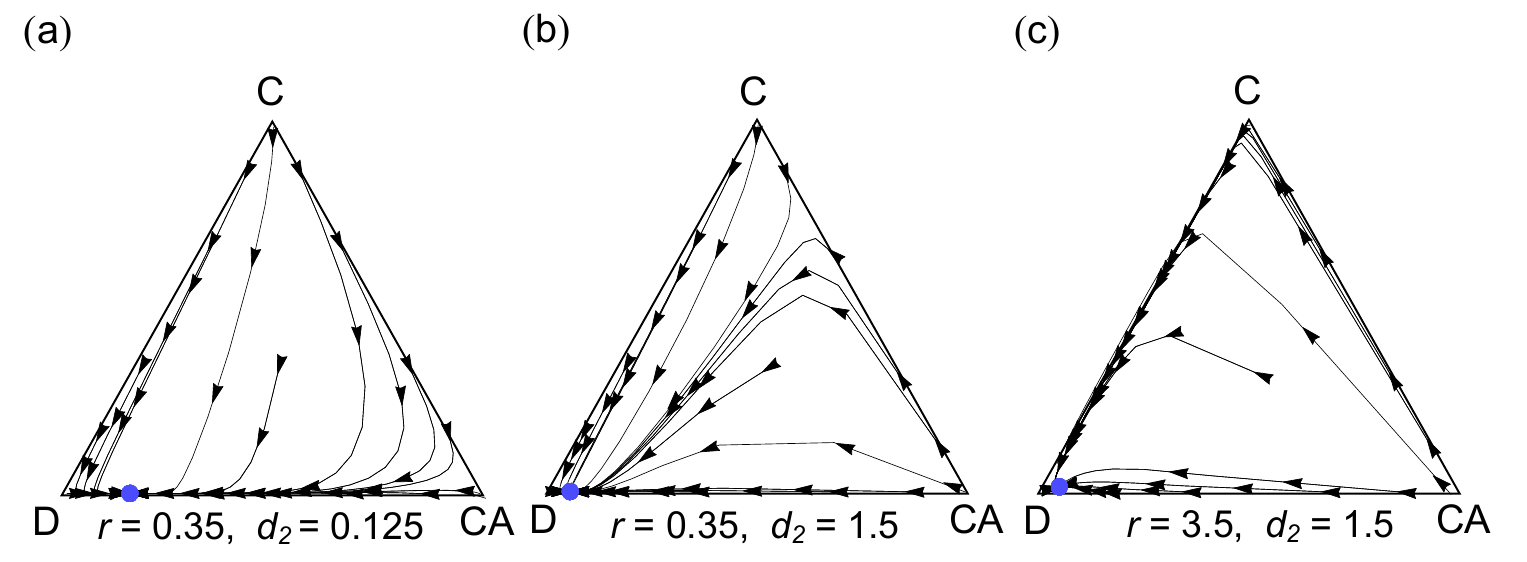}
\caption{\label{fig2}
The evolutionary dynamics of a a large, well-mixed population involving strategies $C$, $D$, and $CA$ are derived from the replicator-mutator dynamics. Each vertex in the plot represents a homogeneous population where all individuals hold the same strategy. The three vertices correspond to homogeneous populations adopting the strategies of cooperation, defection, and constructive agents, respectively. Arrows indicate the direction of dynamics, while blue-purple dots indicate the convergence states of populations. Proximity to a vertex indicates a higher proportion of the strategy represented by that vertex. Figures (a), (b), and (c) respectively depict the evolutionary dynamics in region I (setting $r = 0.35$ and $d_{2}=0.125$), Region II (setting $r=0.35$ and $d_{2}=1.5$), and Region III (setting $r=3.5$ and $d_{2}=1.5$) as shown in Figure~\ref{fig1}. \Reds{The other parameters are set as follows: $N = 5$ and $\mu = 0.005$.} The results presented are generated using a modified version of the DeFinetti package \cite{archetti2020definetti}.
}
\end{figure}

When introducing the constructive agents solely, a similar invasion analysis is conducted under the conditions $0 < r < \Reds{r_{max}=N(M-1)/(M-N)}$ and $d_{2} > 0$, see Appendix A for detailed analysis. The population's invasion state in the presence of constructive agents is entirely different from that introduced by destructive agents. Figure~\ref{fig1} illustrates that the population consistently favors $D$, since $D$ consistently maintain a positive expected payoff greater than the other two types. But the competitive relationship between $C$ and $CA$ is distinct. Region I: when $r < 1-(N-1) d_{2}$ , $C$ can not invade constructive agents, but $CA$ agents can invade $C$, as the expected payoff for the constructive agents is higher than that of cooperators. Region II: When $1-(N-1)d_{2}< r < 1-d_{2}/(M-1)$, $C$ and $CA$ can invade each other, leading to a state of antagonism. Region III: when $r > 1-d_{2}/(M-1)$, $C$ agents can invade $CA$, since the expected payoff of $C$ is greater than that of $CA$. To validate the theoretical analysis, numerical simulations were conducted in three regions, as illustrated in Figure~\ref{fig2}. Consistent with theoretical predictions, across the three mentioned parameter regions, there are variations in the competitive dynamics between $C$ and $CA$, with $D$ consistently holding dominance. These results indicate that when $CA$ exists alone, despite potentially benefiting others, it cannot effectively resolve the social dilemma.

\begin{figure}[!ht]
\centering
\includegraphics[width=1\linewidth]{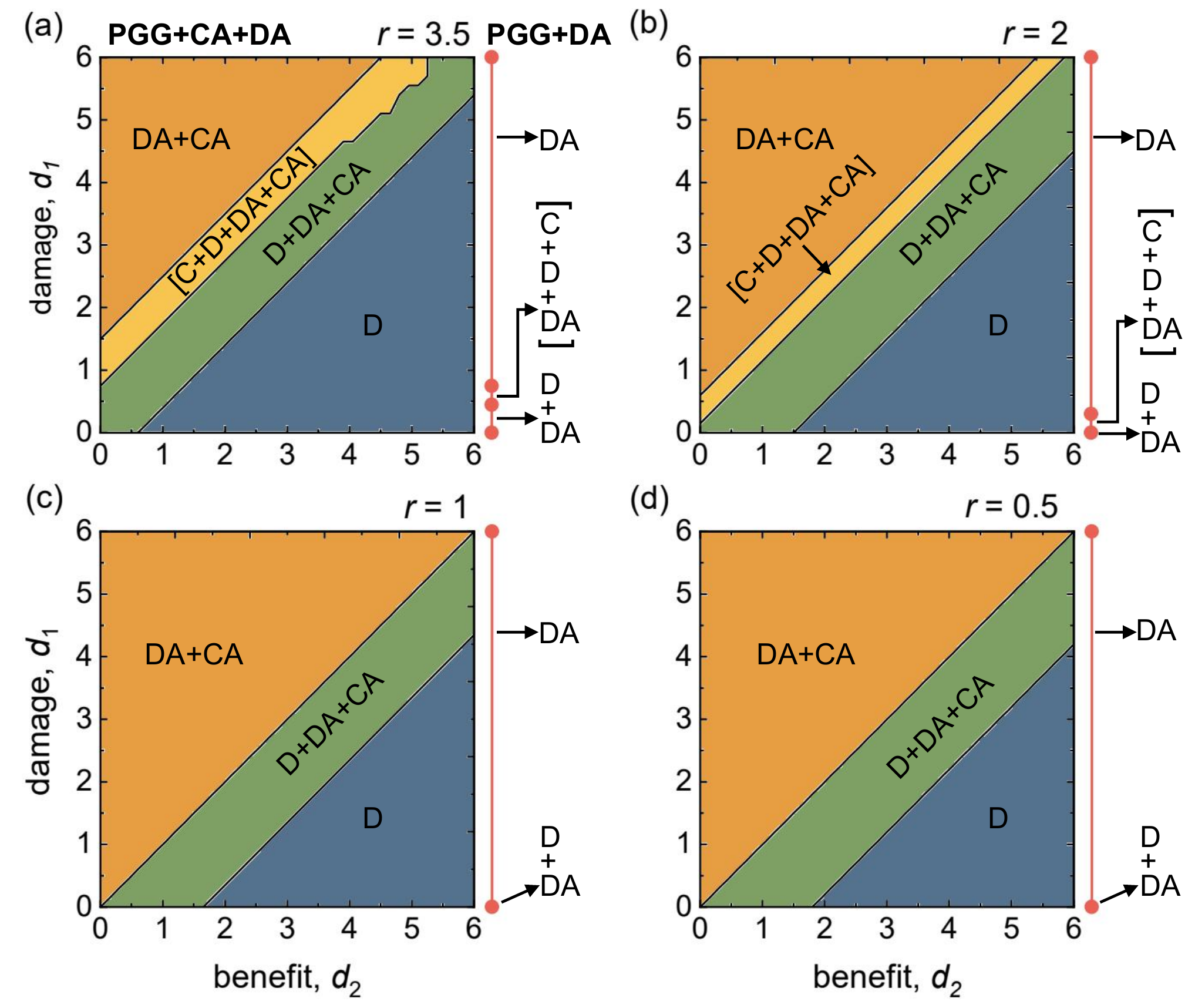}
\caption{\label{fig3}
Depicted are the strategy coexistence in the four-strategy model (involving $C$, $D$, $DA$, and $CA$) as a function of the damage $d_1$ caused by destructive agents and the benefit $d_2$ provided by constructive agents. The red vertical axis on the far right of each subfigure illustrates the coexistence of strategies in the three-strategy model (involving $C$, $D$, and $DA$) under the same parameters. The square brackets denote the presence of cyclic dominance among strategies, while their absence indicates the state of stable coexistence. \Reds{Due to mutations, a given strategy cannot completely disappear or dominate; therefore, a threshold of greater than 0.05 is used to determine its presence.} The values of $r$ are 3.5, 2, 1, and 0.5 for (a)-(d) respectively, with a group size of $N = 5$ and a mutation rate of $\mu = 0.005$.}
\end{figure}

To further explore the impact of constructive agents on cooperation, we investigate the evolutionary dynamics involving the $C$, $D$, $DA$, and $CA$ strategies. The coexistence of strategies obtained through numerical simulation of replicator-mutator equations is depicted in Figure~\ref{fig3}. Both the three-strategy model involving $C$, $D$ and $DA$ strategy and the four-strategy model considered here demonstrate that cooperation occurs only when the synergy factor is sufficiently large (i.e., $r \geq 2$), as shown in Figure~\ref{fig3}(a) and (b). In the four-strategy model, for a specific $d_2$, as the ``harm" inflicted by $DA$ on participants ($d_1$) increases, the population transitions from a pure $D$ (blue region) phase to a coexistence phase of $D+DA+CA$ (green region). Furthermore, as $d_1$ continues to increase, the population shifts from the $D+DA+CA$ phase to the $DA+CA$ (orange region) phase when $r$ is low, as illustrated in Figure~\ref{fig3}(c) and (d). Interestingly, a narrow region of four-strategy coexistence (yellow region) emerges between the $D+DA+CA$ phase and the $DA+CA$ phase when $r$ is sufficiently large. In contrast to the three-strategy model (shown on the right red vertical axis in Figure~\ref{fig3}) where cooperation is limited to low values of $d_1$ \cite{arenas2011joker}, \Reds{the introduction of $CA$ enables cooperation to emerge under higher values of $d_1$. However, the premise for cooperation to be sustained is that $d_1$ must be greater than $d_2$. Once $d_2$ exceeds $d_1$, cooperation diminishes, leading to a transition from the four-strategy coexistence phase to the $D + DA + CA$ phase. Further increases in $d_2$ ultimately lead to the dominance of strategy $D$.}

\begin{figure}[!ht]
\centering
\includegraphics[width=1\linewidth]{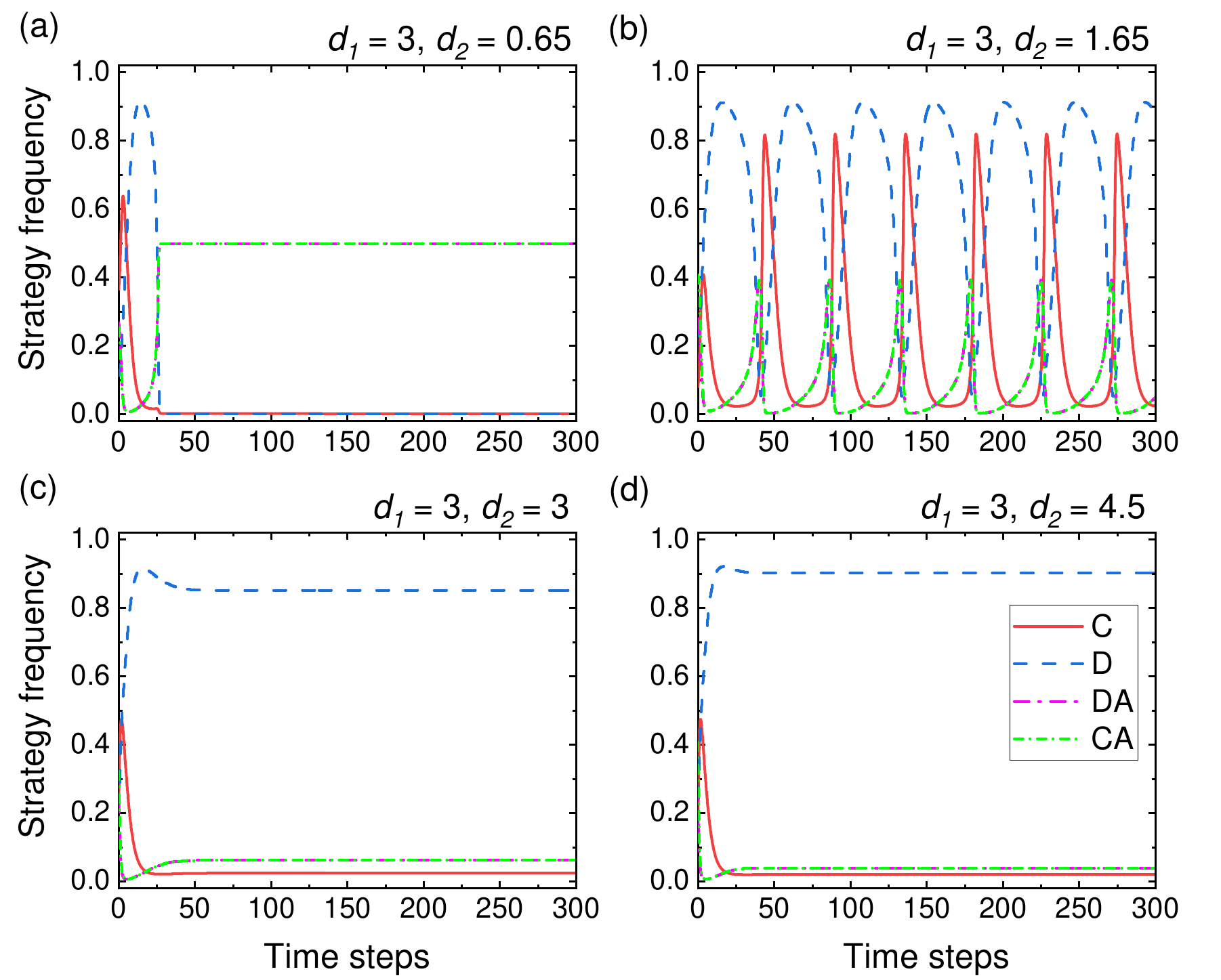}
\caption{\label{fig4} Depicted are the frequency of the $C$ (cooperator, red solid line), $D$ (defector, blue dashed line), $DA$ (destructive agents, magenta dotted line), and $CA$ (constructive agents, green dash-dotted line) as a function of time steps. $d_{2}$ is set to 0.65, 1.65, 3, and 4.5 for (a)-(d), respectively. The other parameters are set as follows: $d_1=3$,  $N = 5$, $r = 3.5$, and $\mu=0.005$. }
\end{figure}

\begin{figure}[!ht]
\centering
\includegraphics[width=1\linewidth]{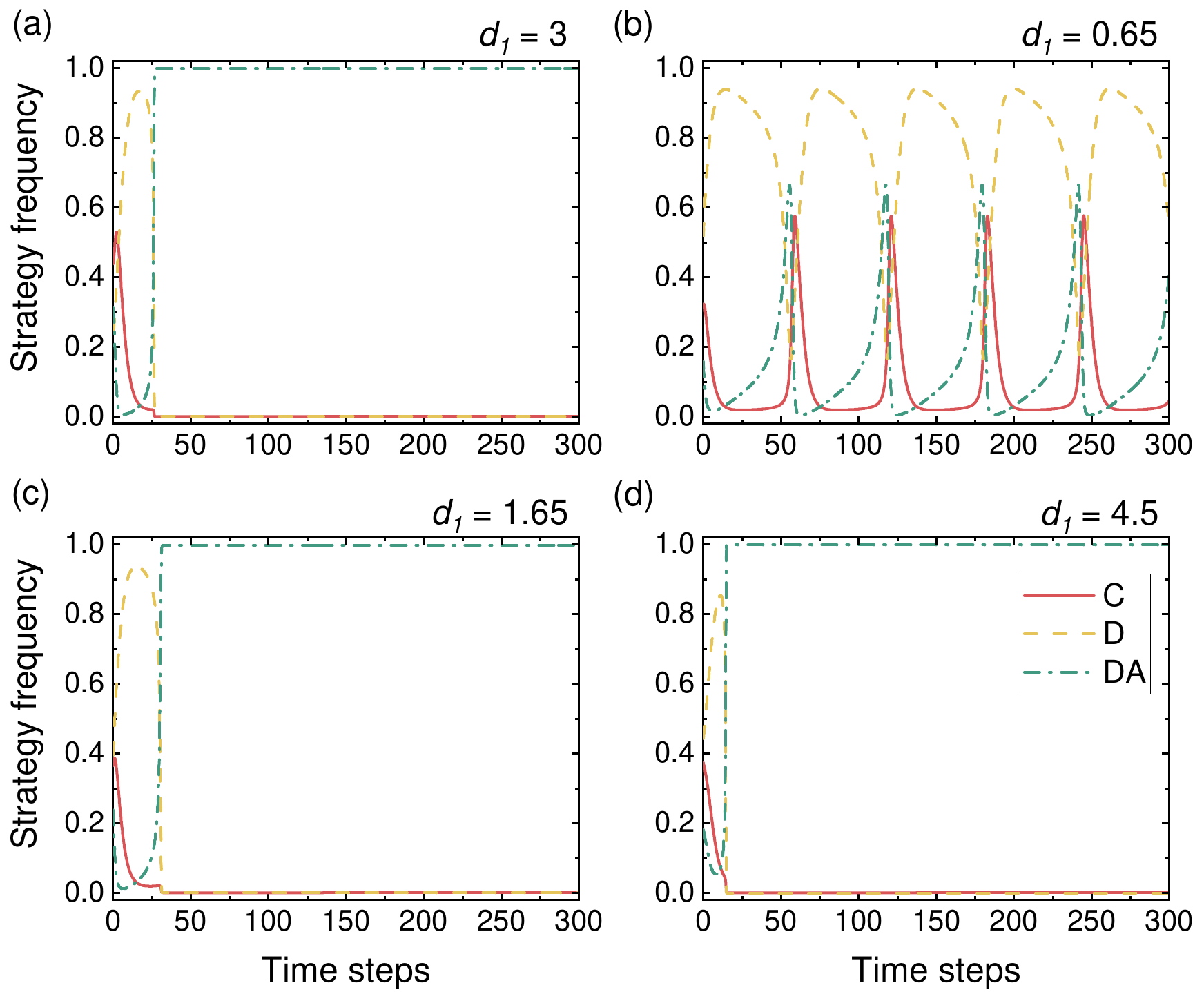}
\caption{\label{fig5}
Depicted are the frequency of the $C$ (cooperator, red solid line), $D$ (defector, yellow dashed line), and $DA$ (destructive agents, green dotted line) as a function of time steps. $d_{1}$ is set to 3, 1.65, 1.65, and 4.5 for (a)-(d), respectively. The other parameters are set as follows:  $N = 5$, $r = 3.5$, and $\mu=0.005$. In (a), the frequency of $C$ briefly increases before giving way to $D$. Strategy $DA$ dominates the entire population eventually. In (b), strategies $C$, $D$, and $DA$ exhibit cyclic dominance. The evolutionary process of strategy proportions in (c) and (d) resemble that of Figure (a), with the eventual dominance of strategy $DA$.
}
\end{figure}

Figure \ref{fig4} illustrates the temporal evolution of four strategies under a synergy factor of $r=3.5$, providing insight into the evolving dynamics within the population. In Figure \ref{fig4}(a), the results indicate a transient increase in the proportions of cooperators and defectors in the early stages of evolution. Subsequently, both $C$ and $D$ decrease while the proportions of $CA$ and $DA$ increase until stability is reached, ultimately resulting in a coexistence state of $CA+DA$. Due to the absence of competition between the $CA$ and $DA$, their proportions in the population are equal at equilibrium. When $d_{2}$ is increased to 1.65, the population exhibits cyclic oscillations over time for these four strategies, as shown in Figure \ref{fig4}(b). Specifically, the proportion of $D$ initially rises, then gives way to $DA$ and $CA$, after which both are gradually replaced by $C$. Then, $C$ is once again supplanted by $D$, forming a cycle. It is this dominating cycle that enables the coexistence of cooperators with the other three strategies. When $d_{1}=d_{2}$, as depicted in Figure \ref{fig4}(c), the dominance of the strategy cycle gives way to the coexistence of $D+DA+CA$. Finally, when the benefits brought by $CA$ outweigh the damage caused by $DA$ (i.e., $d_{2}>d_{1}$), Figure \ref{fig4}(d) shows that defectors eventually dominate the population. In contrast, we examine the temporal evolution dynamics involving three strategies—$C$, $D$, and $DA$. It is noteworthy that when $d_{1} < 1$, the cycle dominance can emerge with $C$, $D$, and $DA$, as shown in Figure \ref{fig5}(b). However, when $d_{1} > 1$, the results in Figures \ref{fig5}(a), (c), and (d) show that cycle dominance disappears and $DA$ ultimately comes to dominate. Overall, \Reds{when both $d_1$ and $d_2$ are larger}, 
the cyclic dominance among strategies can occur (see Figure \ref{fig4}), \Reds{but it must satisfy $d_1 > d_2$}. \Reds{However, as the increasing benefits from $CA$ in the population, the cyclic dominance among strategies induced by destructive agents disappears when $d_2 \geq d_1$},
indicating that the introduction of constructive agents weakens the ability of destructive agents to maintain cooperation within the population. \Reds{The results remain consistent even when constructive and destructive agents are integrated. For more details, refer to Appendix B.}

\begin{figure}[!ht]
\centering
\includegraphics[width=1\linewidth]{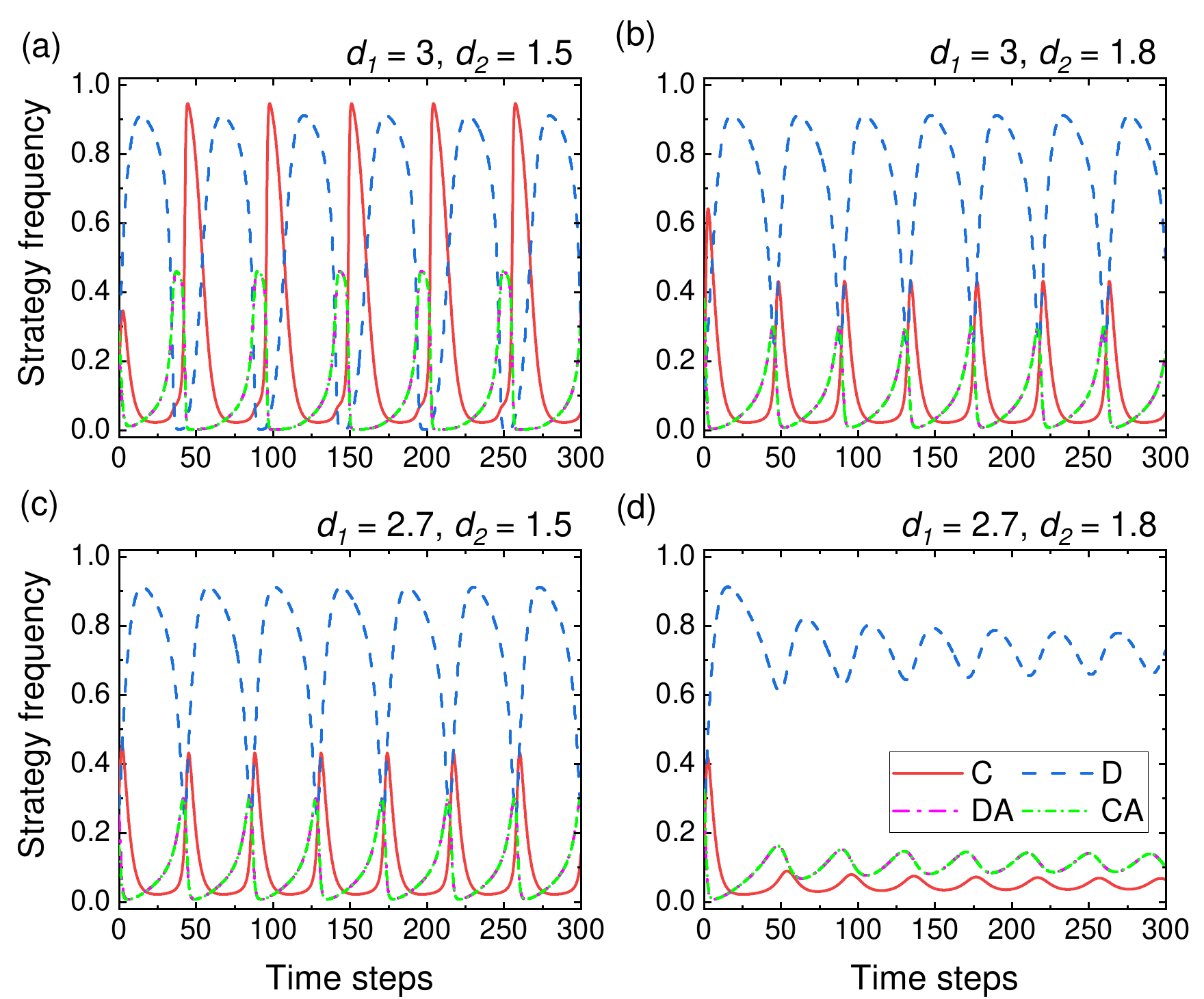}
\caption{\label{fig6}
Depicted are the frequencies of $C$ (cooperator, red solid line), $D$ (defector, blue dashed line), $DA$ (destructive agents, magenta dotted line), and $CA$ (constructive agents, green dash-dotted line) are depicted as a function of time steps under different $d_1$-$d_2$ parameter combinations [$d_1=3$ and $d_2=1.5$ for (a); $d_1=3$ and $d_2=1.8$ for (b); $d_1=2.7$ and $d_2=1.5$ for (c); $d_1=2.7$ and $d_2=1.8$ for (d)]. The other parameters are set as follows: $N = 5$, $r = 3.5$, and $\mu=0.005$. In (a) and (c), a decrease in $d_{1}$ leads to a reduction in the peak frequencies of strategies $C$, $DA$, and $CA$ within the cyclic dominance. In (c) and (d), an increase in $d_{2}$ weakens the peak frequencies of all four strategies within the cyclic dominance.}
\end{figure}

To further investigate the impact of $d_{1}$ and $d_{2}$ on the dynamics of strategy evolution, we examine the coexistence of four strategies ($C$, $D$, $DA$, and $CA$) under varying combinations of $d_{1}$ and $d_{2}$. Results from Figures \ref{fig6}(a) and (b) indicate that, for a fixed $d_{1}$, a slight increase in $d_{2}$ leads to a reduction in the peak amplitude of $C$ during cyclic dominance oscillations (red solid line). Particularly, this weakening trend becomes more pronounced when $d_{1}$ is slightly decreased, as illustrated in Figures \ref{fig6}(c) and (d). This suggests that the increase in benefit ($d_{2}$) brought by the constructive agents weakens the advantage of destructive agents in inducing cooperators to dominate in the population's cyclic dynamics. Decreasing $d_{1}$ also weakens the advantage of cooperators in cyclic dominance.

\begin{figure}[!ht]
\centering
\includegraphics[width=1\linewidth]{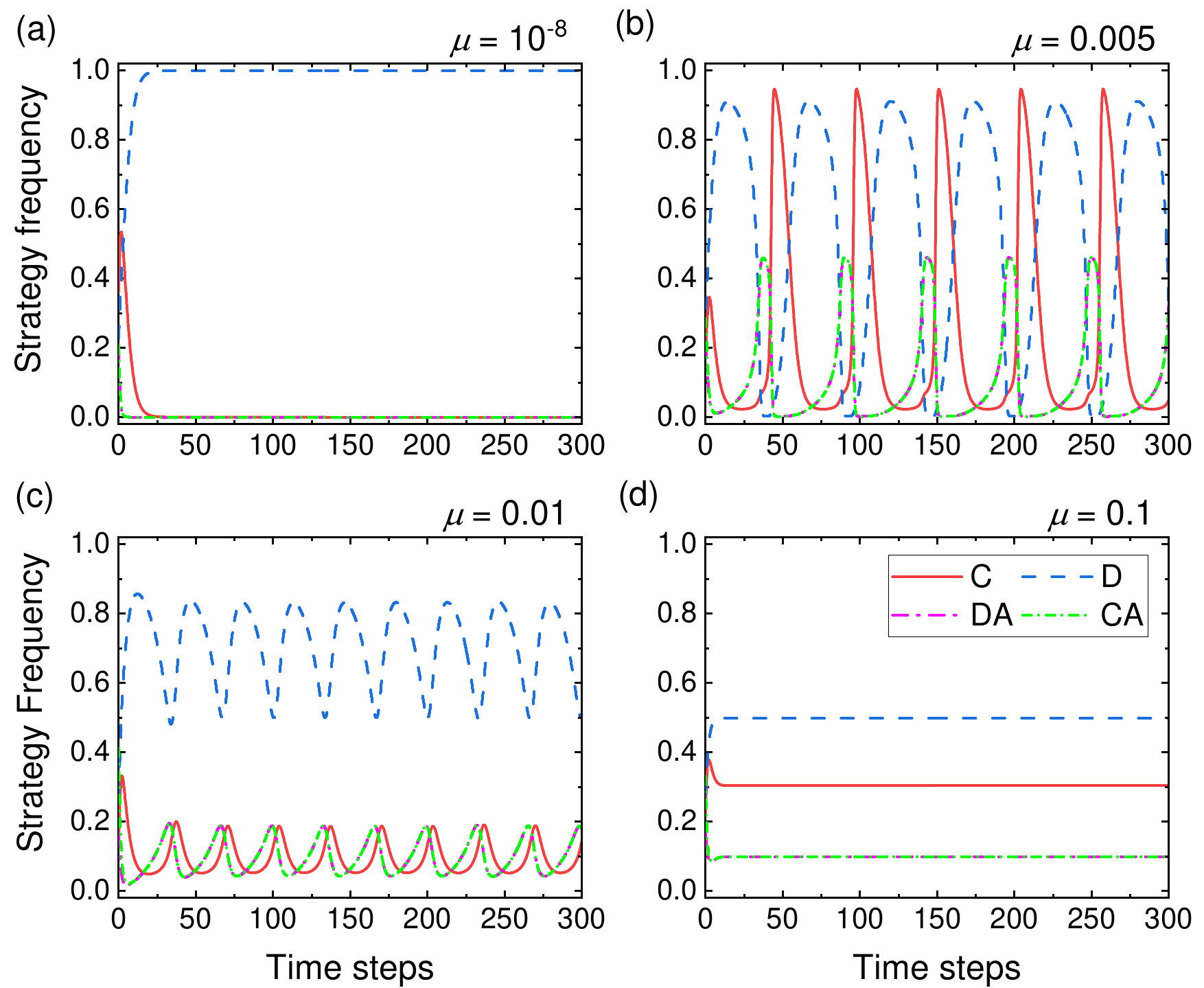}
\caption{\label{fig7}
Depicted are the frequencies of $C$ (cooperator, red solid line), $D$ (defector, blue dashed line), $DA$ (destructive agents, magenta dotted line), and $CA$ (constructive agents, green dash-dotted line) as a function of time steps. The mutation rates are $10^{-8}$, 0.005, 0.01, and 0.1 for (a)-(d) respectively. The other parameters are set as follows: $N = 5$, $r = 3.5$, $d_{1} = 3$, and $d_{2} = 1.5$. As the mutation rate increases, the cyclic dominance among the four strategies gradually diminishes.}
\end{figure}

The impact of mutation rate on the cyclic dominance phenomenon is depicted in Figure \ref{fig7}. When $\mu=0.005$, it can be observed that the peak cooperation frequency within the cyclic dominance reaches approximately 0.92. However, as the mutation rate increases to 0.01, the peak values for the dominance of all four strategies decrease. Specifically, the peak value for cooperators drops from 0.92 to 0.188. When mutation rate increases to 0.1, while all four strategies still coexist within the population, the cyclic dominance phenomenon disappears, with the proportion of defectors surpassing that of cooperators. Moreover, the proportions of $CA$ and $DA$ stabilize at a low level. In the scenario of very rare mutations, specifically $\mu = 10^{-8}$, cyclic dominance disappears entirely, and defectors overwhelmingly dominate, pushing the frequencies of strategies $C$, $CA$, and $DA$ nearly to zero.


\section{Discussion} 

The intertwined evolutionary dynamics of diverse behaviors could deepen understanding of how cooperation is sustained amidst the complexity of human behavior. In this work, we introduce constructive behaviors into the traditional PGG, and consider the intertwined evolutionary dynamics of constructive agents and destructive agents. These constructive agents do not influence contributions and allocations in the public pool. Instead, they directly benefit participants in the PGG group (i.e., cooperators and defectors). Through mean-field results derived from the replicator-mutator equation and numerical simulations, our research reveals significant differences in the impact of constructive agents compared to destructive agents on the cooperation dynamics. While the presence of destructive agents can contribute to sustaining cooperation by facilitating a strategy cycle dominance, the sole participation of constructive agents in the evolutionary dynamics fails to resolve cooperation issues in a one-shot game, with defection still prevailing in the population (see Figure~\ref{fig1}). However, introducing constructive agents into the cooperation-defection-destruction model can maintain cooperation through \Reds{a cyclic dominance process among strategies in evolution (see Figure~\ref{fig3})}.

Although constructive agents themselves cannot directly alter the evolutionary outcomes of the population, together with destructive agents, \Reds{they can engage in a cyclic dominance induced by destructive agents to sustain cooperation (see Figure~\ref{fig3}). Moreover, constructive agents act as a buffer, allowing cooperation to be maintained even when the damage from destructive agents is larger.}
However, as the benefits provided by constructive agents increase, the efficiency of destructive agents in promoting cooperation decreases. When the benefits provided by constructive agents exceed the ``harm'' caused by destructive agents to individuals, it breaks the cyclic dominance and weakens the ability of destructive agents to sustain cooperation (see Figure \ref{fig4}). Furthermore, an increased mutation rate can weaken the peak cooperation frequency in cyclic dominance. Conversely, in the limit of rare mutation rates, the phenomena of cyclic dominance among strategies completely disappear, with the defective strategy being the unique Nash equilibrium. Neither destructive nor constructive agents derive any benefits from a common pool of funds. However, relying solely on altruistic behavior from constructive agents is insufficient to support the establishment of cooperation in one-shot games. Importantly, the altruistic behavior of constructive agents provides positive incentives for others in society, while the behavior of destructive agents generates negative incentives. By simultaneously considering these positive and negative incentive behaviors \cite{balliet2011reward, szolnoki2010reward, szolnoki2017second}, the dynamics of cooperation maintenance are altered. Therefore, our study highlights the importance of balancing the interaction of positive and negative incentive behaviors.

In our study, constructive and destructive agents represent two typical orientations towards the interests of others: benefiting others and causing harm to others. However, individual self-interest considerations have been overlooked. Considering that self-interest is a characteristic of rational human behavior, an integration of individual actions build upon two dimensions of behavioral orientation within the social value orientation framework \cite{bogaert2008social}--preference for one's own gains and preference for opponent's gains--will contribute to a more comprehensive understanding of cooperation emergence and promotion under diverse behavioral frameworks. On the other hand, in this challenging one-shot game scenario, cooperation cannot be sustained by relying solely on the constructive agents. In conjunction with reciprocity mechanisms, can the altruistic behaviors exhibited by constructive agents, promote the development of cooperation? Furthermore, by integrating behavioral information transmission, analyzing the biased strategies of constructive and destructive agents towards defectors or cooperators becomes feasible, aiding in a deeper understanding of how diverse behaviors affect the cooperation dynamics. Future research focusing on direct reciprocity, indirect reciprocity, and network reciprocity mechanisms will offer valuable insights into the complex interplay between diverse individual value orientations and the maintenance of cooperation.

\section*{Article information}
\paragraph*{Acknowledgement} 

We acknowledge the support provided by
(i) the National Natural Science Foundation of China (Grant No.11931015, 12271471, 12161089), Major project of National Philosophy and Social Science Foundation of China (Grants No. 22\&ZD158 and 22VRCO49) to L.S.; (ii) a JSPS Postdoctoral Fellowship Program for Foreign Researchers (Grant No. P21374), and an accompanying Grant-in-Aid for Scientific Research from JSPS KAKENHI (Grant No. JP 22KF0303) to C.S.; (iii) the grant-in-Aid for Scientific Research from JSPS, Japan, KAKENHI (Grant No. JP 20H02314 and JP 23H03499) awarded to J.T; (iv) China Scholarship Council (Grant No.~202308530309) and Yunnan Provincial Department of Education Science Research Fund Project (Grant No. 2024Y503) to Z.H., and (v) Yunnan Provincial Department of Education Science Research Fund Project (Grant No. 2024Y502) to Y.D.

\paragraph*{Author contributions} 
Y.D. and C.S. conceptualised, designed the study, formal analysis, methodology, validation, and visualization; Y.D., Z.H., and C.S.  writing original draft; L.S. and J.T. provided overall project supervision, review and editing; All authors approved the final version and agreed to be accountable for the work conducted in the work.

\paragraph*{Conflict of interest} We declare that no conflict of interests.

\appendix
\renewcommand{\theequation}{A.\arabic{equation}}
\setcounter{equation}{0}
\definecolor{RED}{rgb}{1.0, 0.0, 0.0}

\section*{Appendix A. Finite populations: invasion analysis among cooperation, defection and constructive agents}

Considering a homogeneous population of size $M$, where individuals adopt strategy $Y$. Suppose one individual undergoes a mutation, shifting from strategy $Y$ to strategy $X$. The successful invasion of this homogeneous population by the $X$ individual is contingent upon the condition that the expected payoff from adopting strategy $X$ exceeds that of individuals adopting strategy $Y$ (i.e., $P_{X} > P_{Y}$). In this population, composed of one individual adopting strategy $X$ and $N-1$ individuals adopting strategy $Y$, the expected payoff for the individuals using strategy $X$ can be represented as follows:

\begin{equation}
\begin{gathered}
P_{X}=\pi_{X}(1X, \ (N-1)Y).
\end{gathered}
\label{A.1}
\end{equation}

During the game, individuals adopting strategy $Y$ face two possible population states: one where one individual holds strategy $X$ and $N-1$ individuals hold strategy $Y$, and the other where all individuals hold strategy $Y$. Let's denote the probabilities of these two states as $p_{1}$ and $p_{2}$ respectively. The expected payoff for strategy $Y$ can be expressed as:

\begin{equation}
\begin{gathered}
P_{Y}=\pi_{Y}(1X,(N-1) Y) \cdot p_{1} \\+\pi_{Y}(0X,N Y) \cdot p_{2},
\end{gathered}
\label{A.2}
\end{equation}
where,
\begin{equation}
\begin{split}
p_{1} & = \frac{\binom{M-2}{N-2} \binom{1}{1}}{\binom{M-1}{N-1}} \\
& = \frac{N-1}{M-1},
\end{split}
\label{A.3}
\end{equation}

\begin{equation}
\begin{aligned}
p_{2} & = \frac{\binom{M-2}{N-1} \binom{1}{0}}{\binom{M-1}{N-1}} \\
& = \frac{M-N}{M-1}.
\end{aligned}
\label{A.4}
\end{equation}




In a model involving $C$, $D$, and $CA$ strategies, considering the pairwise interaction of these strategies, there are six distinct scenarios regarding population composition, as outlined below:


\noindent{(A) $1D + (M-1)C$}
\begin{equation}
\begin{aligned}
P_{D} & = \pi_{D}(1 D,(N-1) C) \\
& = r-\frac{r}{N},\\
P_{C} & = \pi_{C}(1 D,(N-1) C) \cdot \frac{N-1}{M-1}+\pi_{C}(0,N C) \cdot \frac{M-N}{M-1} \\
& = r-1-\frac{r}{N} \cdot \frac{N-1}{M-1}.
\end{aligned}
\label{A.5}
\end{equation}

Strategy $D$ will successfully invade the population of strategy $C$, satisfying the $P_{D} > P_{C}$, the condition can be derived as follows: 
\begin{equation}
\begin{aligned}
r<N \cdot \frac{M-1}{M-N}.
\end{aligned}
\label{A.6}
\end{equation}

In the limit $M \to \infty$ , the above condition is reduced to $r < N$, thereby encompassing the condition for the social dilemma: $0 < r < N$. In the subsequent analysis, we will proceed under the assumption that inequality (\ref{A.6}) holds.~\\

\noindent{(B) $1C + (M-1)D$}
\begin{equation}
\begin{aligned}
P_{C} & = \pi_{C}(1 C,(N-1) D)\\
& =\frac{r \cdot 1}{N}-1, \\
P_{D} & = \pi_{D}(1 C,(N-1) D) \cdot \frac{N-1}{M-1}+\pi_{D}(0,N D) \cdot \frac{M-N}{M-1} \\
& = \frac{r \cdot 1}{N} \cdot \frac{N-1}{M-1}.
\end{aligned}
\label{A.7}
\end{equation}

Since $0 < r < N$, we have $P_{C}<0$ and $P_{D}>0$, indicating $P_{C}<P_{D}$. Therefore, strategy $C$ can never invade a population of strategy $D$.~\\

\noindent{(C) $1CA + (M-1)C$}
\begin{equation}
\begin{aligned}
P_{CA} & = 0,\\
P_{C} & = \pi_{C}(1 CA,(N-1) C) \cdot \frac{N-1}{M-1}+\pi_{C}(0,N C) \cdot \frac{M-N}{M-1} \\
& =r-1+\frac{d_{2}}{M-1}.
\end{aligned}
\label{A.8}
\end{equation}

If $P_{CA}>P_{C}$, then:
\begin{equation}
\begin{aligned}
0 < r < 1-\frac{d_{2}}{M-1},
\end{aligned}
\label{A.9}
\end{equation}
strategy $CA$ can invade the population of strategy $C$ iff condition (\ref{A.9}) holds.

On the contrary, let $P_{CA}<P_{C}$ derive : 
\begin{equation}
\begin{aligned}
r>1-\frac{d_{2}}{M-1},
\end{aligned}
\label{A.10}
\end{equation}
when $r$ satisfies condition (\ref{A.10}), the strategy $CA$ can never invade the population of strategy $C$.~\\

\noindent{(D) $1C + (M-1)CA$}
\begin{equation}
\begin{aligned}
P_{CA} & = 0,\\
P_{C} & = \pi_{C}(1 C,(N-1) CA) \\
& =r-1+(N-1) \cdot d_{2}.
\end{aligned}
\label{A.11}
\end{equation}

If $P_{C}>P_{CA}$, thus: 
\begin{equation}
\begin{aligned}
r>1-(N-1) \cdot d_{2},
\end{aligned}
\label{A.12}
\end{equation}
strategy $C$ will invade the population of strategy $CA$ iff condition (\ref{A.12}) holds.

On the contrary, if $P_{C}<P_{CA}$ , i.e.,
\begin{equation}
\begin{aligned}
r<1-(N-1) \cdot d_{2},
\end{aligned}
\label{A.13}
\end{equation}
in this case, the strategy $C$ will not invade the population of strategy $CA$.~\\

\noindent{(E) $1D + (M-1)CA$}
\begin{equation}
\begin{aligned}
P_{CA} & =0,\\
P_{D} & =\pi_{D}(1 D,(N-1) CA) \\
& =d_{2} \cdot(N-1).
\end{aligned}
\label{A.14}
\end{equation}

Since $d_{2} > 0$ and $N > 1$, it follows that $d_{2}(N-1)>0$, i.e., $P_{D}>P_{CA}$. When this condition is satisfied, the  strategy $D$ will inevitably invade the population of strategy $CA$. ~\\

\noindent{(F) $1CA + (M-1)D$}
\begin{equation}
\begin{aligned}
P_{CA} & = 0,\\
P_{D} & = \pi_{D}(1 CA,(N-1) D) \cdot \frac{N-1}{M-1}+\pi_{D}(0,N D) \cdot \frac{M-N}{M-1} \\
& =\frac{d_{2}}{M-1}.
\end{aligned}
\label{A.15}
\end{equation}

Since $d_{2} > 0$ and  $M \gg 1$, it follows that $P_{D}>0$. In this case, $P_{CA}<P_{D}$ must hold, so the strategy $CA$ will never invade the population of strategy $D$.

According to the invasion analysis described above, we have obtained the distinct invasion regions for three strategies as shown in Figure \ref{fig1}. In scenarios (A), (B), (E), and (F), it is evident that when conditions $0 < r < N \cdot \frac{M-1}{M-N}$ and $d_{2} > 0$ are satisfied, both strategy $C$ and strategy $CA$ are consistently invaded by strategy $D$. From scenarios (C) and (D), although $r>1-\frac{d_{2}}{M-1}$ is satisfied, strategy $CA$ does not invade strategy $C$. However, when $r > 1-(N-1) \cdot d_{2}$, strategy $CA$ is invaded by strategy $C$. Notably, the range of $r$ values in the former falls within that of the latter, thus, as long as the condition $r>1-\frac{d_{2}}{M-1}$ holds, the latter will definitely occur, namely, strategy $CA$ being invaded by strategy $C$. Similarly, although $r < 1-(N-1) \cdot d_{2}$ is satisfied, strategy $C$ does not invade strategy $CA$. However, when $0 < r < 1-\frac{d_{2}}{M-N}$, strategy $C$ is invaded by strategy $CA$. The range of $r$ values in the former is contained within the range of the latter, thus, when $r < 1-(N-1) \cdot d_{2}$ is satisfied, strategy $C$ will eventually be invaded by strategy $CA$. In contrast, when $1-(N-1) \cdot d_{2}<r<1-\frac{d_{2}}{M-1}$, strategies $CA$ and $C$ invade each other, leading to a state of antagonism.


\renewcommand{\theequation}{B.\arabic{equation}}
\setcounter{equation}{0}

\section*{Appendix B. Finite populations: invasion analysis of a simplified three-strategy model involving cooperation, defection and integrated exiters }

{The effects that constructive agents and destructive agents have on other group members are opposite, while their own payoffs are zero. According to the replicator-mutator equation (\ref{eq5}), once the proportions of these two types reach equilibrium, the presence of one $CA$ (who increases the payoffs of both cooperators and defectors by $d_{2}$) and one $DA$ (who decreases their payoffs by $d_{1}$) is mathematically equivalent to the presence of one agent who alters the payoffs of cooperators and defectors by $d = d_{2} - d_{1}$. Therefore, we simplify the four-strategy model into a three-strategy model: cooperation ($C$), defection ($D$), and integrated exiters strategies ($E$). Here, each integrated exiter has an effect of $d = d_{2} - d_{1}$, which is equally shared among all cooperators and defectors. In other words, it provides a benefit of $d_{2}$ while also inflicting damage of $d_{1}$ on non-exiters. We conduct a strategy invasion analysis similar to Appendix A for this population of three strategies, examining six distinct combinations through pairwise interactions.}


\noindent{(A) $1D + (M-1)C$}
\begin{equation}
\begin{aligned}
P_{D} & = \pi_{D}(1 D,(N-1) C) \\
& = r-\frac{r}{N},\\
P_{C} & = \pi_{C}(1 D,(N-1) C) \cdot \frac{N-1}{M-1}+\pi_{C}(0,N C) \cdot \frac{M-N}{M-1} \\
& = r-1-\frac{r}{N} \cdot \frac{N-1}{M-1}.
\end{aligned}
\label{B.1}
\end{equation}

If $P_{D} > P_{C}$, strategy $D$ successfully invades the population of strategy $C$ , consistent with (\ref{A.6}), i.e., when
\begin{equation}
\begin{aligned}
r<N \cdot \frac{M-1}{M-N}.
\end{aligned}
\label{B.2}
\end{equation}

\noindent{(B) $1C + (M-1)D$}
\begin{equation}
\begin{aligned}
P_{C} & = \pi_{C}(1 C,(N-1) D)\\
& =\frac{r \cdot 1}{N}-1, \\
P_{D} & = \pi_{D}(1 C,(N-1) D) \cdot \frac{N-1}{M-1}+\pi_{D}(0,N D) \cdot \frac{M-N}{M-1} \\
& = \frac{r \cdot 1}{N} \cdot \frac{N-1}{M-1}.
\end{aligned}
\label{B.3}
\end{equation}

Since  inequality (\ref{B.2}) includes the condition $0 < r < N$, it follows that $P_{C} < P_{D}$, indicating that strategy $C$  will never invade the population of strategy $D$.

\vspace{\baselineskip}
\noindent{(C) $1E + (M-1)C$}
\begin{equation}
\begin{aligned}
P_{E} & = 0,\\
P_{C} & = \pi_{C}(1 E,(N-1) C) \cdot \frac{N-1}{M-1}+\pi_{C}(0,N C) \cdot \frac{M-N}{M-1} \\
& =r-1+\frac{d_{2}-d_{1}}{M-1}.
\end{aligned}
\label{B.4}
\end{equation}

If $P_{E} > P_{C}$, strategy $E$ will invade the population of strategy $C$ , i.e.,
\begin{equation}
\begin{aligned}
0 < r < 1-\frac{d_{2}-d_{1}}{M-1},
\end{aligned}
\label{B.5}
\end{equation}
Conversely, if the following condition is satisfied, i.e.,
\begin{equation}
\begin{aligned}
r>1-\frac{d_{2}-d_{1}}{M-1},
\end{aligned}
\label{B.6}
\end{equation} 
strategy E will not invade the population of strategy C.

\vspace{\baselineskip}
\noindent{(D) $1C + (M-1)E$}
\begin{equation}
\begin{aligned}
P_{E} & = 0,\\
P_{C} & = \pi_{C}(1 C,(N-1) E) \\
& =r-1+(N-1) \cdot (d_{2}-d_{1}).
\end{aligned}
\label{B.7}
\end{equation}

If $P_{C} > P_{E}$, then strategy $C$ will invade the population of strategy $E$, i.e., 
\begin{equation}
\begin{aligned}
r>1-(N-1) \cdot (d_{2}-d_{1}),
\end{aligned}
\label{B.8}
\end{equation}
Conversely, if $P_{C} < P_{E}$, strategy $C$ will not invade the population of strategy $E$, i.e.,
\begin{equation}
\begin{aligned}
r<1-(N-1) \cdot (d_{2}-d_{1}),
\end{aligned}
\label{B.9}
\end{equation}

\noindent{(E) $1D + (M-1)E$}
\begin{equation}
\begin{aligned}
P_{E} & =0,\\
P_{D} & =\pi_{D}(1 D,(N-1) E) \\
& = (d_{2}-d_{1}) \cdot(N-1).
\end{aligned}
\label{B.10}
\end{equation}

In the above equation (\ref{B.10}), if $d_{2} - d_{1} > 0$, then $P_{D} > P_{E}$, and strategy $D$ will invade the population of strategy $E$. Conversely, if $d_{2} - d_{1} < 0$, then $P_{D} < P_{E}$, and strategy $D$ does not invade the population of strategy $E$. When $d_{2} = d_{1}$,  strategy $D$ and strategy $E$ cannot invade each other.

\vspace{\baselineskip}
\noindent{(F) $1E + (M-1)D$}
\begin{equation}
\begin{aligned}
P_{E} & = 0,\\
P_{D} & = \pi_{D}(1 E,(N-1) D) \cdot \frac{N-1}{M-1}+\pi_{D}(0,N D) \cdot \frac{M-N}{M-1} \\
& =\frac{d_{2}-d_{1}}{M-1}.
\end{aligned}
\label{B.11}
\end{equation}

Similarly, in equation (\ref{B.11}), if $d_{2} - d_{1} < 0$, then $P_{E} > P_{D}$, and strategy $E$ will invade the population of strategy $D$. Conversely, if $d_{2} - d_{1} > 0$, then $P_E < P_D$, and strategy $E$ does not invade the population of strategy $D$. When $d_{2} = d_{1}$, strategy $D$ and strategy $E$ cannot invade each other.

The above strategy invasion analysis demonstrates that the evolutionary dynamics of the population depend on the relative magnitudes of $d_{2}$ and $d_{1}$. When $d_{2} - d_{1} < 0$, a $ C \rightarrow D \rightarrow E \rightarrow C $ cycle dominance emerges under the condition $ r>1-(N-1) \cdot (d_{2}-d_{1}) $. For $1-\frac{d_{2}-d_{1}}{M-1} < r < 1-(N-1) \cdot (d_{2}-d_{1})$, strategy $C$ and strategy $E$ cannot invade each other, resulting in a bistable state. When $0< r < 1-\frac{d_{2}-d_{1}}{M-1}$, strategy $E$ invades both the populations of strategy $C$ and strategy $D$, ultimately leading to the dominance of strategy $E$. \Reds{Thus, a population consisting of strategies $C$, $D$, and $E$ can sustain cooperation through a cyclic dominance process, primarily attributed to the destructive effect caused by the integrated exiters, i.e., when $d_{2}-d_{1} < 0$.}

Conversely, when $d_{2} - d_{1} > 0$ , if $r < 1-(N-1) \cdot (d_{2}-d_{1})$, strategy $E$ will invade the population of strategy $C$, and both strategies $E$ and $C$ will be invaded by strategy $D$. If $ 1-(N-1) \cdot (d_{2}-d_{1}) < r < 1-\frac{d_{2}-d_{1}}{M-1}$, strategy $E$ and strategy $C$ will invade each other, but both will be invaded by strategy $D$. If $r > 1-\frac{d_{2}-d_{1}}{M-1}$, strategy $C$ will invade the population of strategy $E$, and strategy $D$ will invade both strategy $C$ and strategy $E$.

However, when $d_{2}-d_{1} = 0$, it follows from (\ref{B.10}) and (\ref{B.11}) that $PE = PD = 0$, indicating that strategies $D$ and $E$ cannot invade each other, and there is no competition between them. For strategies C and E, according to (\ref{B.4}) and (\ref{B.7}), we have $P_{C} = r-1$ and $P_{E} = 0$. Therefore, when $r > 1$, strategy $E$ does not invade the population of strategy $C$; instead, strategy $C$ invades the population of strategy $E$. Additionally, based on (\ref{B.1}) and (\ref{B.3}), strategy $D$ always invades the population of strategy $C$. Thus, this leads to $E \rightarrow C \rightarrow D$, resulting in the dominance of strategy $D$. Conversely, when $r < 1$, strategy $C$ does not invade strategy $E$, while strategy $E$ invades strategy $C$. Combined with (\ref{B.1}) and (\ref{B.3}), since strategy $D$ always invades strategy $C$, both strategies $E$ and $D$ invade strategy $C$, ultimately leading to the coexistence of strategies $E$ and $D$.

The evolutionary dynamics involving cooperators, defectors, and integrated exiters in the simplified three-strategy model depend on $d = d_{2} - d_{1}$. Cooperation can still be sustained through a three-strategy cyclic dominance involving cooperators, defectors, and integrated exiters when $ d_{2} - d_{1} < 0 $, indicating the critical role of destructive agents in maintaining cooperation. \Reds{However, when $d_{2} - d_{1} > 0$, the presence of integrated exiters does not alter the equilibrium of defection. In other words, when the benefits provided by constructive agents surpass the damage caused by destructive agents, constructive agents nullify the ability of destructive agents to maintain cooperation.}

\bibliographystyle{unsrt}
\bibliography{sample}

\begin{thebibliography}{10}

\bibitem{axelrod1981evolution}
Robert Axelrod and William~D Hamilton.
\newblock The evolution of cooperation.
\newblock {\em science}, 211(4489):1390--1396, 1981.

\bibitem{riolo2001evolution}
Rick~L Riolo, Michael~D Cohen, and Robert Axelrod.
\newblock Evolution of cooperation without reciprocity.
\newblock {\em Nature}, 414(6862):441--443, 2001.

\bibitem{archetti2012game}
Marco Archetti and Istvan Scheuring.
\newblock Game theory of public goods in one-shot social dilemmas without assortment.
\newblock {\em Journal of theoretical biology}, 299:9--20, 2012.

\bibitem{weibull1997evolutionary}
J{\"o}rgen~W Weibull.
\newblock {\em Evolutionary game theory}.
\newblock MIT press, 1997.

\bibitem{nowak2011supercooperators}
Martin Nowak and Roger Highfield.
\newblock {\em Supercooperators: Altruism, evolution, and why we need each other to succeed}.
\newblock Simon and Schuster, 2011.

\bibitem{rand2013human}
David~G Rand and Martin~A Nowak.
\newblock Human cooperation.
\newblock {\em Trends in cognitive sciences}, 17(8):413--425, 2013.

\bibitem{guo2024cooperation}
Hao Guo, Zhen Wang, Junliang Xing, Pin Tao, and Yuanchun Shi.
\newblock Cooperation and coordination in heterogeneous populations with interaction diversity.
\newblock In {\em Proceedings of the 23rd International Conference on Autonomous Agents and Multiagent Systems}, pages 752--760, 2024.

\bibitem{nowak2006five}
Martin~A Nowak.
\newblock Five rules for the evolution of cooperation.
\newblock {\em science}, 314(5805):1560--1563, 2006.

\bibitem{schmid2021unified}
Laura Schmid, Krishnendu Chatterjee, Christian Hilbe, and Martin~A Nowak.
\newblock A unified framework of direct and indirect reciprocity.
\newblock {\em Nature Human Behaviour}, 5(10):1292--1302, 2021.

\bibitem{leimar2001evolution}
Olof Leimar and Peter Hammerstein.
\newblock Evolution of cooperation through indirect reciprocity.
\newblock {\em Proceedings of the Royal Society of London. Series B: Biological Sciences}, 268(1468):745--753, 2001.

\bibitem{fehr2000cooperation}
Ernst Fehr and Simon G{\"a}chter.
\newblock Cooperation and punishment in public goods experiments.
\newblock {\em American Economic Review}, 90(4):980--994, 2000.

\bibitem{balliet2011reward}
Daniel Balliet, Laetitia~B Mulder, and Paul~AM Van~Lange.
\newblock Reward, punishment, and cooperation: a meta-analysis.
\newblock {\em Psychological bulletin}, 137(4):594, 2011.

\bibitem{szolnoki2010reward}
Attila Szolnoki and Matjaz Perc.
\newblock Reward and cooperation in the spatial public goods game.
\newblock {\em Europhysics Letters}, 92(3):38003, 2010.

\bibitem{li2018punishment}
Xuelong Li, Marko Jusup, Zhen Wang, Huijia Li, Lei Shi, Boris Podobnik, H~Eugene Stanley, Shlomo Havlin, and Stefano Boccaletti.
\newblock Punishment diminishes the benefits of network reciprocity in social dilemma experiments.
\newblock {\em Proceedings of the National Academy of Sciences}, 115(1):30--35, 2018.

\bibitem{szolnoki2017second}
Attila Szolnoki and Matja{\v{z}} Perc.
\newblock Second-order free-riding on antisocial punishment restores the effectiveness of prosocial punishment.
\newblock {\em Physical Review X}, 7(4):041027, 2017.

\bibitem{andreoni2003carrot}
James Andreoni, William Harbaugh, and Lise Vesterlund.
\newblock The carrot or the stick: Rewards, punishments, and cooperation.
\newblock {\em American Economic Review}, 93(3):893--902, 2003.

\bibitem{chen2015first}
Xiaojie Chen, Tatsuya Sasaki, {\AA}ke Br{\"a}nnstr{\"o}m, and Ulf Dieckmann.
\newblock First carrot, then stick: how the adaptive hybridization of incentives promotes cooperation.
\newblock {\em Journal of the royal society interface}, 12(102):20140935, 2015.

\bibitem{szabo2002phase}
Gy{\"o}rgy Szab{\'o} and Christoph Hauert.
\newblock Phase transitions and volunteering in spatial public goods games.
\newblock {\em Physical review letters}, 89(11):118101, 2002.

\bibitem{szabo2002evolutionary}
Gy{\"o}rgy Szab{\'o} and Christoph Hauert.
\newblock Evolutionary prisoner’s dilemma games with voluntary participation.
\newblock {\em Physical Review E}, 66(6):062903, 2002.

\bibitem{shen2021exit}
Chen Shen, Marko Jusup, Lei Shi, Zhen Wang, Matja{\v{z}} Perc, and Petter Holme.
\newblock Exit rights open complex pathways to cooperation.
\newblock {\em Journal of the Royal Society Interface}, 18(174):20200777, 2021.

\bibitem{li2024granting}
Shulan Li, Zhixue He, Danyang Jia, Chen Shen, Lei Shi, and Jun Tanimoto.
\newblock Granting leaders priority exit options promotes and jeopardizes cooperation in social dilemmas.
\newblock {\em Neurocomputing}, page 127566, 2024.

\bibitem{bogaert2008social}
Sandy Bogaert, Christophe Boone, and Carolyn Declerck.
\newblock Social value orientation and cooperation in social dilemmas: A review and conceptual model.
\newblock {\em British journal of social psychology}, 47(3):453--480, 2008.

\bibitem{murphy2011measuring}
Ryan~O Murphy, Kurt~A Ackermann, and Michel~JJ Handgraaf.
\newblock Measuring social value orientation.
\newblock {\em Judgment and Decision making}, 6(8):771--781, 2011.

\bibitem{arenas2011joker}
Alex Arenas, Juan Camacho, Jos{\'e}~A Cuesta, and Rub{\'e}n~J Requejo.
\newblock The joker effect: Cooperation driven by destructive agents.
\newblock {\em Journal of theoretical biology}, 279(1):113--119, 2011.

\bibitem{requejo2012stability}
Rub{\'e}n~J Requejo, Juan Camacho, Jos{\'e}~A Cuesta, and Alex Arenas.
\newblock Stability and robustness analysis of cooperation cycles driven by destructive agents in finite populations.
\newblock {\em Physical Review E}, 86(2):026105, 2012.

\bibitem{khatun2024stability}
Khadija Khatun, Chen Shen, Lei Shi, and Jun Tanimoto.
\newblock Stability of pairwise social dilemma games: destructive agents, constructive agents, and their joint effects.
\newblock {\em arXiv preprint arXiv:2402.12809}, 2024.

\bibitem{grund2013natural}
Thomas Grund, Christian Waloszek, and Dirk Helbing.
\newblock How natural selection can create both self-and other-regarding preferences and networked minds.
\newblock {\em Scientific reports}, 3(1):1480, 2013.

\bibitem{choi2013strategic}
Jung-Kyoo Choi and TK~Ahn.
\newblock Strategic reward and altruistic punishment support cooperation in a public goods game experiment.
\newblock {\em Journal of Economic Psychology}, 35:17--30, 2013.

\bibitem{gneezy2012conflict}
Ayelet Gneezy and Daniel~MT Fessler.
\newblock Conflict, sticks and carrots: war increases prosocial punishments and rewards.
\newblock {\em Proceedings of the Royal Society B: Biological Sciences}, 279(1727):219--223, 2012.

\bibitem{santos2012role}
Francisco~C Santos, Flavio~L Pinheiro, Tom Lenaerts, and Jorge~M Pacheco.
\newblock The role of diversity in the evolution of cooperation.
\newblock {\em Journal of theoretical biology}, 299:88--96, 2012.

\bibitem{perc2008social}
Matja{\v{z}} Perc and Attila Szolnoki.
\newblock Social diversity and promotion of cooperation in the spatial prisoner’s dilemma game.
\newblock {\em Physical Review E}, 77(1):011904, 2008.

\bibitem{smith1982evolution}
John~Maynard Smith.
\newblock Evolution and the theory of games.
\newblock In {\em Did Darwin get it right? Essays on games, sex and evolution}, pages 202--215. Springer, 1982.

\bibitem{archetti2020definetti}
Marco Archetti.
\newblock Definetti: a mathematica program to analyze the replicator dynamics of 3-strategy collective interactions.
\newblock {\em SoftwareX}, 11:100415, 2020.

\end{thebibliography}

\end{document}